\documentclass[graybox]{svmult}

\usepackage{type1cm}        
\usepackage{makeidx}         
\usepackage{graphicx}        
\usepackage{multicol}        
\usepackage[bottom]{footmisc}

\usepackage{newtxtext}       %
\usepackage[varvw]{newtxmath}       

\makeindex             

\begin{document}

\title*{SAIC: Integration of Speech Anonymization and Identity Classification}
\author{Ming Cheng\orcidID{0000-0002-6422-1748} \and Xingjian Diao\orcidID{0000-0001-9605-4991} \and Shitong Cheng\orcidID{0009-0008-3830-7880} \and Wenjun Liu\orcidID{0009-0002-3953-1661}}
\institute{
The first two authors are joint first authors. \\
Ming Cheng \and  Xingjian Diao \and  Shitong Cheng \and  Wenjun Liu \at Department of Computer Science, Dartmouth College, 15 Thayer Drive, Hanover, NH 03755, USA \\
Ming Cheng \\
\email{ming.cheng.gr@dartmouth.edu}\\
Xingjian Diao \\
\email{xingjian.diao.gr@dartmouth.edu} \\
Shitong Cheng \\
\email{shitong.cheng.gr@dartmouth.edu} \\
Wenjun Liu \\
\email{wenjun.liu.gr@dartmouth.edu}
}

%
%
\maketitle

\abstract*{Speech anonymization and de-identification have garnered significant attention recently, especially in the healthcare area including telehealth consultations, patient voiceprint matching, and patient real-time monitoring. Speaker identity classification tasks, which involve recognizing specific speakers from audio to learn identity features, are crucial for de-identification. Since rare studies have effectively combined speech anonymization with identity classification, we propose SAIC -- an innovative pipeline for integrating \textbf{S}peech \textbf{A}nonymization and \textbf{I}dentity \textbf{C}lassification. SAIC demonstrates remarkable performance and reaches state-of-the-art in the speaker identity classification task on the Voxceleb1 dataset, with a top-1 accuracy of $96.1\%$. Although SAIC is not trained or evaluated specifically on clinical data, the result strongly proves the model's effectiveness and the possibility to generalize into the healthcare area, providing insightful guidance for future work.}

\abstract{Speech anonymization and de-identification have garnered significant attention recently, especially in the healthcare area including telehealth consultations, patient voiceprint matching, and patient real-time monitoring. Speaker identity classification tasks, which involve recognizing specific speakers from audio to learn identity features, are crucial for de-identification. Since rare studies have effectively combined speech anonymization with identity classification, we propose SAIC -- an innovative pipeline for integrating \textbf{S}peech \textbf{A}nonymization and \textbf{I}dentity \textbf{C}lassification. SAIC demonstrates remarkable performance and reaches state-of-the-art in the speaker identity classification task on the Voxceleb1 dataset, with a top-1 accuracy of $96.1\%$. Although SAIC is not trained or evaluated specifically on clinical data, the result strongly proves the model's effectiveness and the possibility to generalize into the healthcare area, providing insightful guidance for future work.}

\section{Introduction}
\label{sec:1}
Significant research has been focused on using AI techniques for anonymization and de-identification in the ethics and healthcare area, especially for health record and patient notes protection \cite{zuccon2014identification, ahmed2020identification, dernoncourt2017identification, venugopal2022privacy}.
Meanwhile, the anonymization focusing on speech has not been widely explored, with several studies only developing methods on a limited scale of datasets \cite{han2020voice, chen2023voicecloak}. In parallel, speaker identity classification tasks, which require accurately identifying individuals from their audio \cite{audiomae, niizumi2022masked}, play a crucial role in privacy protection services. These tasks involve disentangling the unique vocal characteristics  (voiceprint) of a person, essentially understanding the speaker's identity information within speech. While this precision is valuable in itself, it also opens up possibilities for enhancing speech anonymization techniques. Ideally, if a system can understand and isolate identity features in speech, it could then modify, obscure, or remove these features to anonymize the audio effectively. Since limited work has implemented the integration of anonymization and identity classification, there remains an unsolved challenge: \textit{Is it feasible to develop a model that simultaneously achieves high-quality speech anonymization and maintains accurate speaker identity classification?}


To address such a research gap, we propose SAIC -- a novel pipeline for speech anonymization and identity classification. After training, SAIC can extract accurate content/identity embeddings, removing identity information from the original audio.  Moreover, it can merge the content of one audio with the voiceprint of another speaker, generating a synthesized speech that maintains content integrity with an altered identity.

In summary, our contribution is threefold:
\begin{itemize}
    \item {
    We propose \textbf{SAIC}, a novel pipeline \textbf{integrating speech anonymization and identity classification} effectively. The content embeddings and identity embeddings are extracted with high quality through the robust encoders. 
    }
    \item { 
     The identity classification task on the Voxceleb1 dataset outperforms existing work and results in the \textbf{state-of-the-art}, with a top-1 accuracy of $96.1\%$. 
    }
    \item {
    SAIC is capable of \textbf{synthesizing new audio} by merging the content from one speaker's audio with the voiceprint of another, effectively generating a synthesized speech that preserves the original content while adopting a different vocal identity.  
    }
\end{itemize}

\section{Related Work}
\subsection{Speech Anonymization and De-Identification}
Initially, a novel approach known as DROPSY \cite{justin2015speaker} is proposed to conceal the speaker's identity. It builds a diphone recognition system for speech recognition, followed by a speech synthesis system to transform a speaker's speech into that of a different individual. In a separate endeavor, VoicePrivacy \cite{tomashenko2020introducing} is proposed to propel advancements in speech data anonymization. The benchmark established by it seeks to minimize the disclosure of the speaker's identity while preserving the distinctiveness of the speech.

Recent studies have emerged in the domain of voice privacy preservation, proposing methods that span multi-dimensional aspects, such as differentially private approach \cite{shamsabadi2022differentially},
naturalness and timbre-preserving \cite{deng2023v}, 
and adversarial examples \cite{chen2023voicecloak}. However, they use a limited size of validation dataset and mainly focus on specific scenarios. This potentially limits the generalizability of the proposed techniques. 

To address the research gap mentioned above, we propose SAIC, a novel pipeline for speaker de-identification and privacy preservation. We evaluate our model on a commonly used and large-scale dataset, Voxceleb1 \cite{nagrani2017voxceleb}, with state-of-the-art results indicating the effectiveness of our model. 

\subsection{Speaker Identity Classification}
The task of speaker identity classification has garnered significant attention in recent years, driven by its applications in various domains including privacy protection, voice-controlled systems, and human-computer interaction. 

With the development of Transformers \cite{vaswani2017attention} and ViT \cite{dosovitskiy2020image}, multiple studies have adopted these architectures as the backbone. For example, SS-AST \cite{gong2022ssast}
pretrains the AST model with joint discriminative and generative masked spectrogram patch modeling, while 
wav2vec 2.0 \cite{baevski2020wav2vec} focuses on learning powerful representations from speech
audio. Although ViT-based methods outperform CNN-based ones in various AI tasks, they usually require massive data and repeated pretraining, failing to handle temporal dynamics without strong data augmentations \cite{islam2022recent, he2022masked}. Therefore, our model is conducted with CNN as the backbone and follows the mainstream encoder-decoder structure \cite{lu2016training, toshniwal2017multitask, hu2020dasgil, karita2018sequence}, achieving significant results and requiring fewer computational resources.

Considering the effectiveness of MAE-based methods on various downstream tasks \cite{he2022masked, gong2022contrastive, audiomae, diao2023av, tong2022videomae}, recent work mainly follows the mask-and-reconstruction strategy for representation learning to do identity classification \cite{maeast, niizumi2022masked, m2d}. However, these methods do not consider the integration of identity classification and audio anonymization and lack strong identity disentanglement capabilities. Based on this, we propose SAIC that can effectively remove identity information while also achieving superior classification performance.

\section{Method}

We address the challenge of implementing de-identification for speaker privacy protection through the proposed SAIC pipeline. 
Formally, given the input audio of speaker $i$ and speaker $j$, our goal is to synthesize the new audio with the speaker $j$'s identity and speaker $i$'s content information. This aims to remove the identity information from the speaker $i$ for privacy protection. 

\begin{figure*}[h]
\centering
\includegraphics[width=\textwidth]{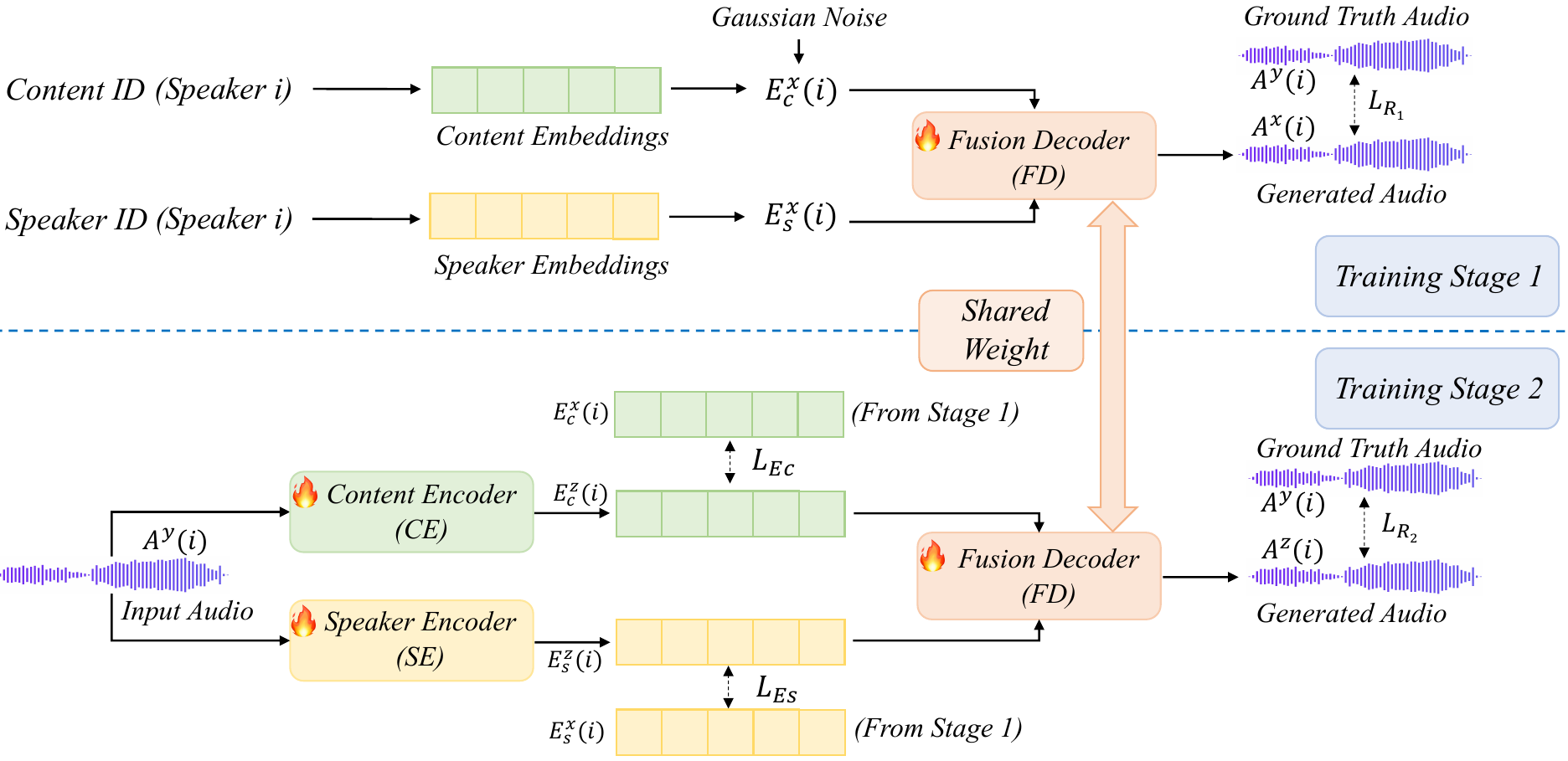}
   \caption{\textbf{SAIC pipeline training.}
   The flame icon indicates trainable modules.
   Inspired by \cite{gabbay2019demystifying}, 
   the pipeline training includes 2 stages: The first stage utilizes latent optimization to obtain accurate content and speaker embeddings and optimize the Fusion Decoder (\textit{FD}) through audio reconstruction. During the second stage, 
   the original audio is input into the Content Encoder (\textit{CE}) and Speaker Encoder (\textit{SE}) to extract specific embeddings, followed by a Fusion Decoder (\textit{FD}, with the same weight as stage 1) to reconstruct the audio.
   }
\label{fig:SAIC}
\end{figure*}

\begin{figure*}[h]
\centering
\includegraphics[width=\textwidth]{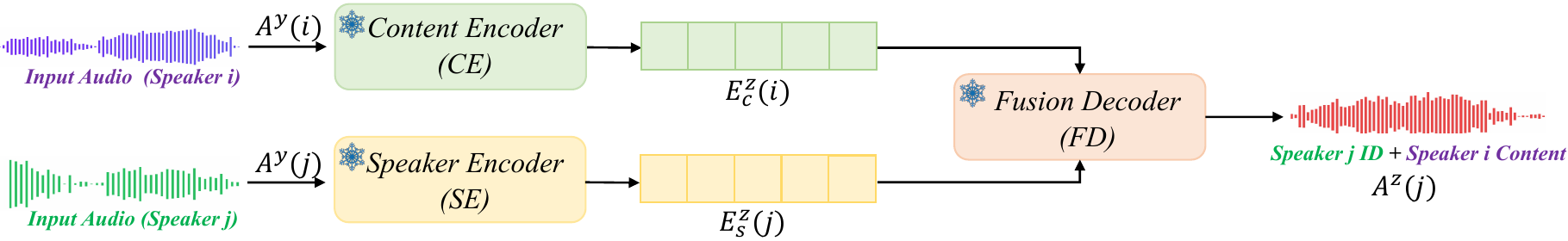}
   \caption{\textbf{SAIC pipeline inference.}
   The snowflake icon represents frozen parameters. 
   During inference, two audio ($A^y(i), A^y(j)$) sourced from different speakers are input into the well-trained Content Encoder (\textit{CE}) and Speaker Encoder (\textit{SE}), respectively. The synthesized audio contains no identity information of the speaker $i$.
   }
\label{fig:SAIC_inf}
\end{figure*}

The training and inference of the SAIC pipeline are shown in Figure \ref{fig:SAIC} and \ref{fig:SAIC_inf}, respectively. 
Inspired by \cite{gabbay2019demystifying}, the pipeline training contains 2 stages. The first stage aims to extract accurate content embeddings ($E_c^x(i) \sim x$) and speaker embeddings ($E_s^x(i) \sim x$) from content and speaker ID, where $x$ indicates latent space. Moreover, the Fusion Decoder (\textit{FD}) is trained to reconstruct the original audio through the latent optimization strategy \cite{gabbay2019demystifying}. 
The second stage focuses on optimizing the Content Encoder (\textit{CE}), Speaker Encoder (\textit{SE}), and the Fusion Decoder (\textit{FD}) to reconstruct audio.  
Specifically, 
suppose $A^y(i) \sim y$ as the input audio of speaker $i$ where $y$ indicates the ground truth domain, it is input into \textit{CE} and \textit{SE} to extract content embeddings ($E_c^z(i) \sim z$) and speaker identity embeddings ($E_s^z(i)  \sim z$), respectively, where $z$ indicates the latent space from two encoders. 
Afterward, $E_c^z(i)$ and $E_s^z(i)$ are input into \textit{FD} to reconstruct audio. Through this pipeline, the two encoders and the decoder can be well-trained for inference. 

During inference, we apply audio input from two different speakers $i, j$, aiming to remove the speaker $i's$ identity information. In this phase, all encoders and the decoder are frozen.

\subsection{Two-Stage Pipeline Training}

\subsubsection{Training Stage 1}
The first stage aims to extract accurate content and speaker embeddings through latent optimization \cite{gabbay2019demystifying}. Specifically, given content and speaker ID, the corresponding embeddings, $E_c^x(i)$ and $E_s^x(i)$, are obtained in the latent space. The two embeddings are then input into the Fusion Decoder (\textit{FD}) to reconstruct the audio. To train \textit{FD} through latent optimization strategy, we employ VGG perceptual loss \cite{hoshen2019non} as $L_{R_1}$ for each speaker $i$:
\begin{equation}
    \begin{split}
        L_{R_1} &= \sum_i^n ||A^x(i) - A^y(i)|| + \lambda||\epsilon_i||^2, \quad \epsilon_i \sim N(0, \sigma^2I) \\
        A^x(i) &= FD(E_s^x(i), E_c^x(i) + \epsilon_i)
    \end{split}
\end{equation}
where $FD$ is the Fusion Decoder, $n$ indicates all speakers, and  $\epsilon_i$ represents the Gaussian Noise of fixed variance and an active attenuation penalty embedding which is applied to content embeddings ($E_c^x(i)$) to regularize the content. 

After stage 1, accurate content and speaker embeddings can be obtained and the decoder is well-trained to generate audio from two embeddings. 

\subsubsection{Training Stage 2}
In stage 2 of pipeline training, we design an encoder-decoder-based architecture to generate the audio. Specifically, 
we construct two encoders ($CE$ and $SE$) to learn accurate embeddings from original audio, and use $FD$ with shared weight from the first stage to reconstruct the audio. 

To instruct SAIC to learn the exact content and speaker embeddings, we employ MSE loss as the embedding loss: 
\begin{equation}
    L_{E_c} = \left\|E_{c}^x(i) - E_{c}^z(i)\right\|^2, \quad
    L_{E_s} = \left\|E_{s}^x(i) - E_{s}^z(i)\right\|^2
\end{equation}
where $E_{c}^x(i)$ and $E_{s}^x(i)$ are the embeddings obtained from the first stage. 

Similar to the first stage, we apply VGG perceptual loss as the reconstruction loss $L_{R_2}$ (without Gaussian noise) to guide the model to generate correct and precise audio:
\begin{equation}
    \begin{split}
        L_{R_2} &= \sum_i^n ||FD(E_s^z(i), E_c^z(i)) - A^y(i)||
    \end{split}
\end{equation}

Finally, the combined loss for pipeline training in the second stage is expressed below:
\begin{equation}
    L_{total} = \lambda_1L_{E_c} + \lambda_2L_{E_s} + \lambda_3L_{R_2}
\end{equation}

\begin{figure*}[h]
\begin{center}
\includegraphics[width=\textwidth]{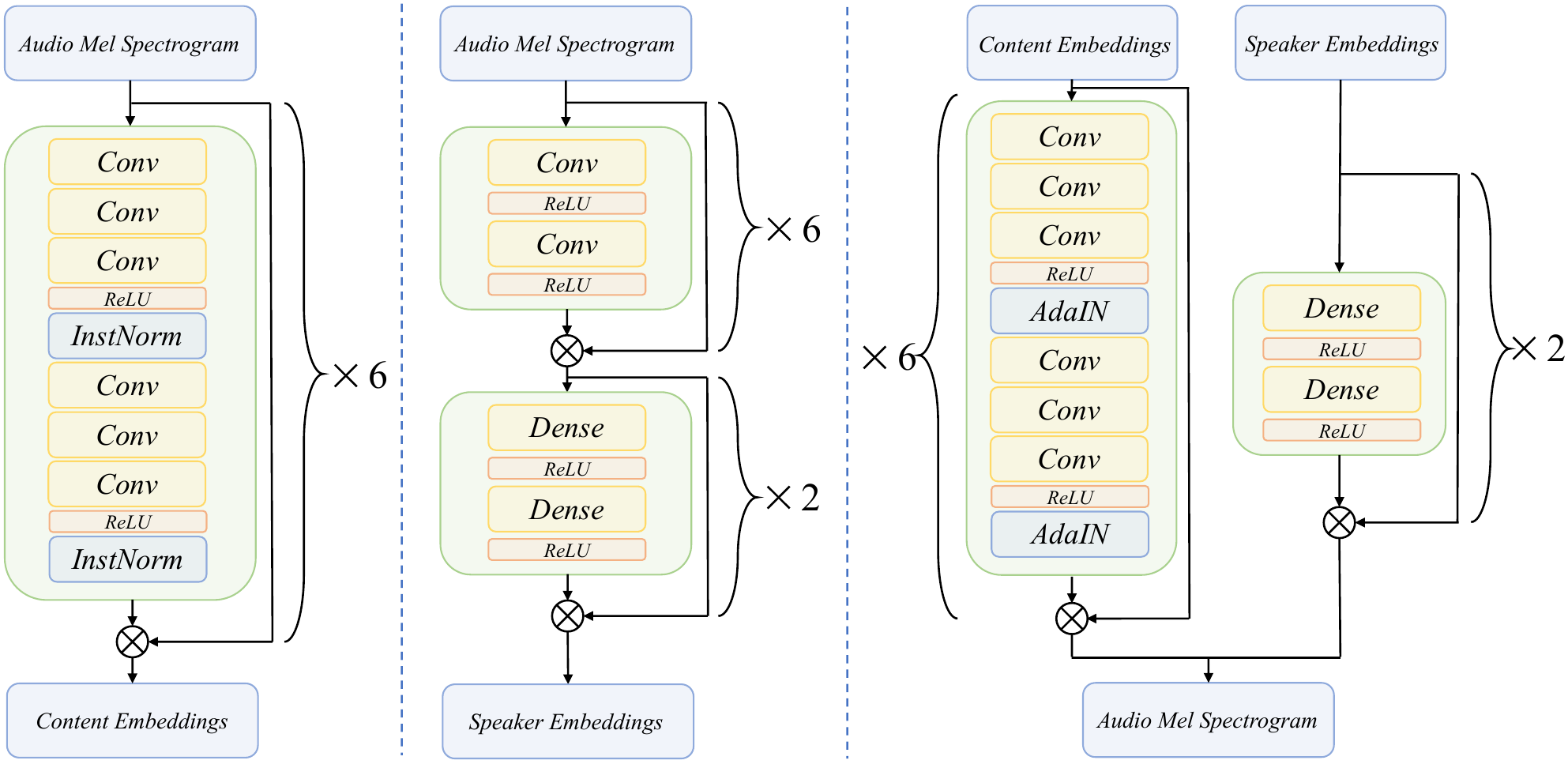}
\end{center}
   \caption{\textbf{Detailed architecture of trainable modules.} \textbf{Left: }Content Encoder. \textbf{Middle:} Speaker Encoder.\textbf{ Right:} Fusion Decoder. 
   }
\label{fig:encoderdecoder}
\end{figure*}

\subsection{Content Encoder}
 The Content Encoder (\textit{CE}) is constructed through sequential convolution blocks. Following \cite{chou2019one}, it uses residual blocks repeated 6 times as the main structure. Each residual block includes a convolutional module and an instance normalization \cite{huang2017arbitrary}. This structure aims to extract the high-dimensional content information in audio to omit the identity information in the content embedding.

\subsection{Speaker Encoder}
The Speaker Encoder (\textit{SE}) contains a sequential residual block (repeated 6 times) and a sequential fully connected dense layer (repeated twice), following \cite{chou2019one}. Similar to the design of \textit{CE}, the residual blocks aim to extract high-dimensional identity information from the audio, followed by the fully connected dense layer to map the extracted features into the specified dimension and fully split the embedding of different speakers. 

\subsection{Fusion Decoder}
The Fusion Decoder (\textit{FD}) fuses features of content/identity embeddings for audio generation. 
Specifically, it consists of two sub-modules, with the first one 
decoding content embeddings through sequential convolution blocks and the second one decoding voice print features through dense layers, based on \cite{chou2019one}. AdaIN (Adaptive Instance Normalization) \cite{huang2017arbitrary} is applied in the decoder. 







\section{Experiments}

We conduct experiments on the VoxCeleb1 dataset \cite{nagrani2017voxceleb} and evaluate the de-identification results through the identity classification task. 

\subsection{Dataset}
The VoxCeleb1 dataset is a diverse audio dataset in the real environment collected from public YouTube videos. It contains 153,516 audio clips from 1,251 speakers with various ages, roles, identities, \textit{etc.} This dataset can be used for speech recognition, speaker classification, and speech information analysis. To evaluate the de-identification quality of our model, we conduct the speaker classification task and compare the results with other related works.

\subsection{Evaluation Metrics}
The evaluation of our model follows the steps below:
\begin{itemize}
    \item {
    For each speaker $i$ in the test set, we randomly choose a different speaker $j$ as the input of \textit{CE} and \textit{SE}. 
    }
    \item {
    After SAIC, the synthesized audio $A^z(j)$ contains the identity information of speaker $j$ and speech content from speaker $i$. $A^z(j)$ is then input into a powerful pre-trained \textit{VoiceEncoder} \cite{wan2018generalized}  from the \textit{Resemblyzer} library in Python for speaker identity embeddings extraction.
    }
    \item {
    Finally, we utilize the extracted embeddings by \textit{VoiceEncoder} to find the corresponding speaker ID. Since we only find one best-matched speaker each time, we report 
    the top-1 accuracy as the evaluation metric.
    }
\end{itemize}
\subsection{Quantitative Results}
\subsubsection{Comparison With State-of-The-Art}

\begin{table*}[h]
\caption{
\textbf{Comparison with state-of-the-art on speaker identity classification task on Voxceleb1 dataset.}
Different from previous methods in which Transformer \cite{vaswani2017attention} is used and mask-and-reconstruction strategy (MAE) \cite{he2022masked} is applied, SAIC is implemented using a CNN-based architecture without MAE strategy and achieves the highest accuracy.  
}
\centering
\resizebox{0.7\textwidth}{!}{%
\begin{tabular}{c|ccc}
\hline\noalign{\smallskip}
\textbf{Methods}     & \textbf{Backbone} & \textbf{MAE-Based}  & \textbf{Top1-Acc} \\ 
\noalign{\smallskip}\svhline\noalign{\smallskip}
\textbf{MAE-AST \cite{maeast}}     & ViT-B    &\checkmark        & 63.3              \\
\textbf{SS-AST \cite{gong2022ssast}}      & ViT-B     & $\times$        & 64.3              \\
\textbf{wav2vec 2.0 \cite{baevski2020wav2vec}} & Transformer
& $\times$ & 75.2              \\
\textbf{HuBERT \cite{hsu2021hubert}}      & Transformer  & $\times$      & 81.4              \\
\textbf{M2D \cite{m2d}}         & ViT-B     &\checkmark        & 94.8              \\
\textbf{Audio-MAE \cite{audiomae}}   & ViT-B    &\checkmark         & 94.8              \\
\textbf{MSM-MAE \cite{niizumi2022masked}}     & ViT-B    &\checkmark         & 95.3              \\ 
\noalign{\smallskip}\svhline\noalign{\smallskip}
\textbf{SAIC (Ours)} & CNN     &   $\times$        & \textbf{96.1}     \\ 
\noalign{\smallskip}\hline\noalign{\smallskip}
\end{tabular}
}
\label{bigtable}
\end{table*}

The quantitative result of our model compared with other related work is shown in Table \ref{bigtable}.
From Table \ref{bigtable},
we can observe that our model provides the state-of-the-art speaker identity classification result, with a top-1 accuracy of $96.1\%$. 

Since the latest work is mainly MAE-based structures that follow the mask-and-reconstruction strategy, this result proves the effectiveness of our disentanglement approach, with a lead of top-1 accuracy of $0.8\%$ \cite{m2d} and $1.3\%$ \cite{audiomae}, respectively. 

In addition to the effectiveness of speaker identity classification, our model has application-level advantages over the MAE-based ones. Since SAIC disentangles the speaker identity and the content of the speech with high quality (only about 3.9\% of the speaker identities are not fully extracted), our model can be applied in many privacy-preserving real-world applications in the future, especially in the healthcare area: telehealth consultations, patient voiceprint matching, patient real-time monitoring, \textit{etc.}

\subsubsection{Transformers vs. CNN}
As from Table \ref{bigtable}, recently, the most relevant work utilizes either Transformers \cite{vaswani2017attention} or ViT \cite{dosovitskiy2020image} as the backbone, whereas SAIC chooses ResBlocks as the main body of the architecture. Considering the fact that Transformer-based models mainly require massive data and strong data augmentations to handle temporal dynamics and variations \cite{islam2022recent, diao2023ft2tf}, our CNN-based architecture showcases the advantages of model efficiency. The ResBlock's capability to effectively handle local features and temporal dynamics makes it more suitable for tasks requiring granular audio analysis and sustained temporal coherence, such as in speech disentanglement and synthesis, where accurate extraction and reconstruction of audio across time are crucial for achieving high-quality outcomes.

Moreover, our model achieves a significant lead compared with some Transformer-based ones ($+20.9\%$ against wav2vec 2.0 \cite{baevski2020wav2vec}, $+14.7\%$ against HuBERT \cite{hsu2021hubert}). This showcases the effectiveness of CNN over Transformers on small datasets or pipeline training without strong data augmentations, having significant advantages in the healthcare area where obtaining large-scale datasets is rarely possible.

\subsection{Qualitative Results}

\begin{figure*}[h]
\centering
\includegraphics[width=1.0\textwidth]{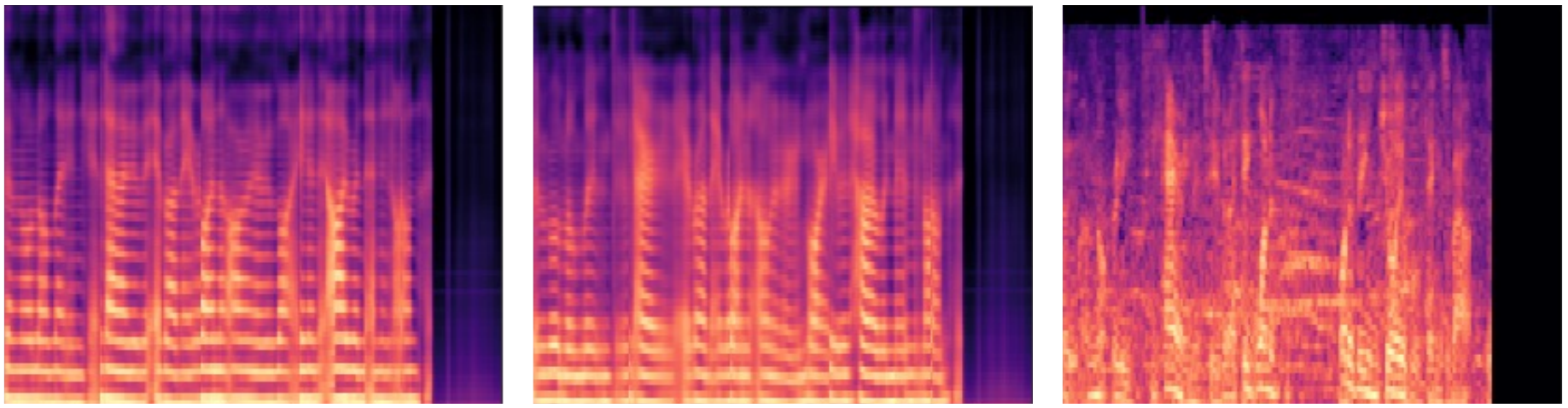}
\caption{\textbf{Visualization of the audio mel spectrogram.} 
\textbf{Left:} Original audio. 
\textbf{Middle: } Reconstructed audio.
\textbf{Right: } Synthesized audio (different identity).
SAIC can reconstruct high-quality audio during training and can anonymize the identity during inference.
}
\label{fig:vis}
\end{figure*}

\begin{figure}[h]
    \centering
    \includegraphics[width=\linewidth]{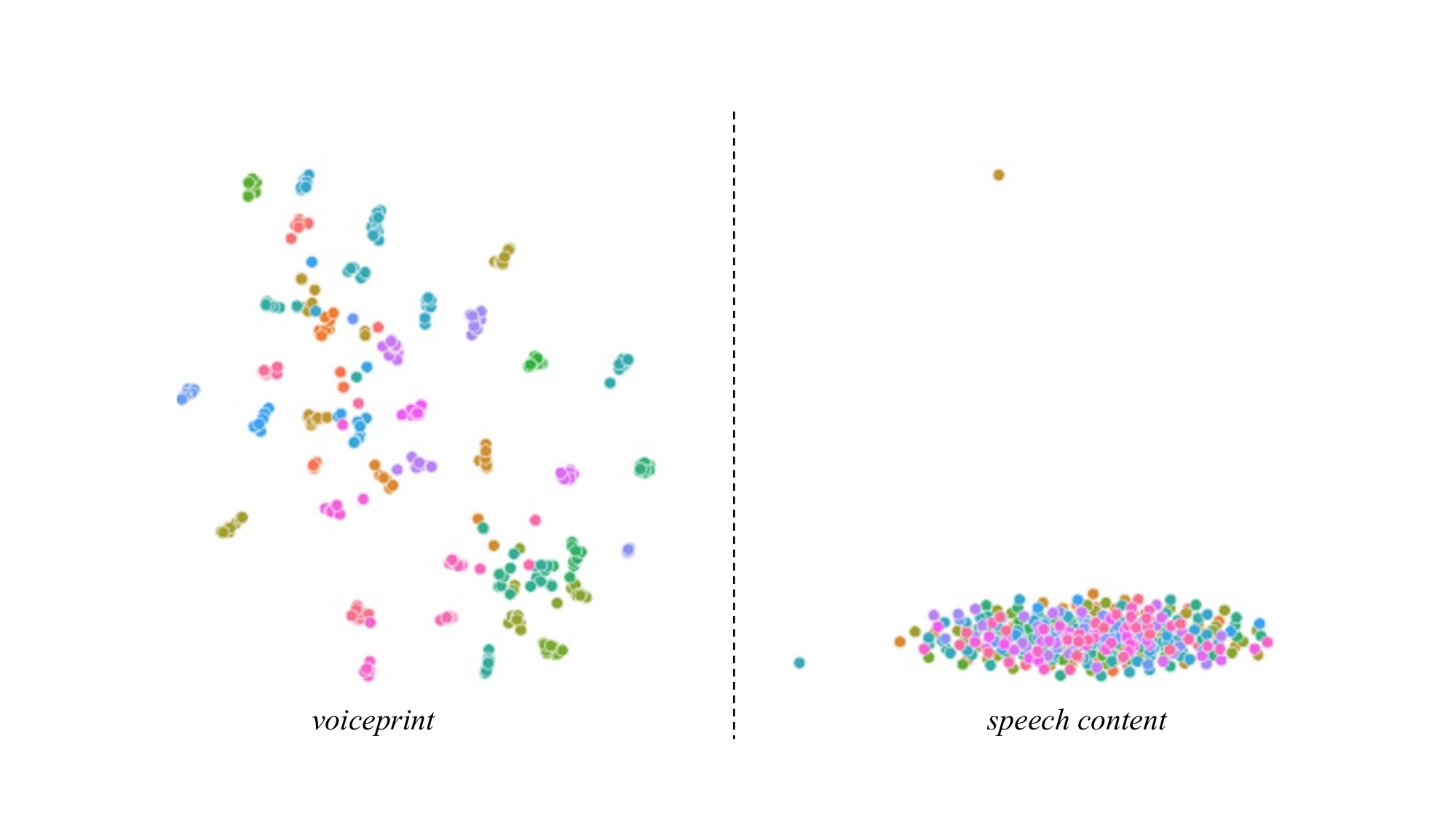}
    \caption{
    \textbf{t-SNE visualization for different speakers on Voxceleb1.} Results on 40 speakers are shown as an example. 
    \textbf{Left:}  Voiceprint results. \textbf{Right:} Content embeddings results.
    Each particle indicates features from a distinct speech. 
    }
    \label{fig:tsne}
\end{figure}

\subsubsection{Audio Reconstruction and Synthesis}
The visualization of audio reconstruction during training and audio generation during inference is shown in Figure \ref{fig:vis}.
Since the original mel spectrogram and the reconstructed one are highly similar, we can evidently observe that SAIC can reconstruct accurate and precise audio through the disentanglement strategy. This indicates that all the trainable modules (\textit{CE}, \textit{SE}, \textit{FD}) are optimized well. Furthermore, during inference, our model utilizes the identity of a different speaker, resulting in a distinctly different mel spectrogram. This clearly demonstrates the model's ability to de-identify audio effectively, thus providing robust privacy protection.
As for real-world applications, our model can be deployed for 
secure handling of sensitive patient information during telehealth consultations, medical dictations, and other audio-based interactions.

\subsubsection{Audio Disentanglement}

The qualitative result of audio disentanglement by t-SNE on the Voxceleb1 dataset is shown in Figure \ref{fig:tsne}, where each particle indicates features from a distinct speech (Results on 40 speakers are shown as an example). 
As for voiceprint disentanglement, 
SAIC can distinguish each speaker clearly, which is reflected in the figure that particles of the same color are clustered together, while there is a distance between particles of different colors. 
Meanwhile, it is clear to observe that the content embeddings of the
audio of different speakers are all clustered into one blob. Since the t-SNE two-dimensional mapping cannot cluster such complex information as audio semantics,
this shows that the content embedding disentangled by the
Content Encoder is exactly audio content information. Also, the content embedding information contains
almost no identity information of the speaker. 
This visualization result showcases that SAIC can be further utilized for healthcare privacy protection.

\section{Conclusion}
In this paper, we introduce SAIC, a novel pipeline integrating speech disentanglement and speaker classification effectively. 
SAIC is constructed with CNN as the backbone containing an encoder-decoder architecture. 
The well-trained model is utilized for speech de-identification and new audio generation during the inference phase. 
The identity classification results on the Voxceleb1 dataset highly prove the effectiveness of SAIC, with top-1 accuracy of $96.1\%$.

Although SAIC is not trained and evaluated specifically on clinical data, the result strongly proves the model’s effectiveness and possibility to generalize into the healthcare area, including patient voiceprint matching, real-time monitoring, \textit{etc.}

\begin{acknowledgement}
We express our gratitude to Dartmouth College alumnus Gokul Srinivasan and Professor SouYoung Jin from the Department of Computer Science, Dartmouth College, Hanover, US, for their invaluable support and contributions throughout our research process.
\end{acknowledgement}

\end{document}